\documentclass{aastex63}
\usepackage{amsmath}
\usepackage{xcolor}

\begin{document}

\title{Resolution Study of Thermonuclear Initiation in White Dwarf Tidal Disruption Events}

\author{Peter Anninos}
\affil{Lawrence Livermore National Laboratory, Livermore, CA 94550, USA}

\author{Karen D. Camarda}
\affil{Department of Physics and Astronomy, Washburn University, Topeka, KS 66621, USA}

\author{Brooke Estes-Myers}
\affil{Department of Physics and Astronomy, Washburn University, Topeka, KS 66621, USA}

\author{Nathaniel Roth}
\affil{Lawrence Livermore National Laboratory, Livermore, CA 94550, USA}

\begin{abstract}
We study the initiation of thermonuclear detonations in tidally disrupted white dwarf stars
by intermediate-mass ($10^3 M_\odot$) black holes. The length scales required to resolve the initiation
mechanism are not easily reached in three-dimensions, so instead
we have devised two-dimensional proxy models which, together with a logarithmic
gridding strategy, can adequately capture detonation wave fronts as material undergoes
simultaneous compression and stretching from tidal forces. We consider 0.15 and 0.6 solar mass
white dwarf stars parameterized by tidal strengths in the range $\beta=4~\text{to}~23$.
High spatial resolution elucidates the manner and conditions leading to thermonuclear
detonation, linking the initiation sequence to stellar composition and tidal strength.
All of our models suffer sustained detonations triggered by a combination of adiabatic compression,
mild thermonuclear preconditioning, and collisional heating, in degrees depending primarily on tidal strength.
We find many diagnostics, such as temperature, total released energy, and iron group products,
are fairly well-converged (better than 10\%) at resolutions below 10 km along the scale height of the 
orbital plane. The exceptions are intermediate
mass transients like calcium, which  remain uncertain up to factors of two even at 1 km resolution.
\end{abstract}


\section{Introduction}
\label{sec:intro}

Tidal disruptions (TD) of white dwarf (WD) stars by intermediate mass black holes (BH, IMBH)
create hot dense environments capable of initiating thermonuclear detonations
and producing isotopically rich material composed of both intermediate and heavy (iron group) nuclei
\citep{Luminet89, Wilson04, Rosswog09, Tanikawa17, Kawana18, Anninos18, Anninos19}.
The compressive forces exerted by near encounters with $10^3$ - $10^4$ solar mass black holes drive
white dwarf cores to temperatures and densities in excess of
$10^9$ Kelvin and $10^7$ g cm$^{-3}$, where helium and carbon/oxygen burning become efficient
relative to hydrodynamic relaxation timescales.
Burn products, if synthesized in sufficient quantity and dispersed as unbound debris, give
rise to observable electromagnetic transients through the decay and reprocessing of unstable isotopes \citep{Macleod16, Kawana20}.
The observational implications of these transients are far-reaching, as the data may provide
evidence for the existence of IMBHs and an estimate of their mass function 
\citep{Gerssen02, Gerssen03, Gebhardt02, Gebhardt05, Dong07}.
They are also potential sources of sub-luminous light curves generally associated with supernovae.
Observations continue to motivate the development of ever more accurate and detailed
numerical reactive models in an effort to better understand tidal disruption events
and their characteristic emissions.

Whether (and to what degree) thermonuclear burn occurs depends strongly on the tidal strength
($\beta=R_T/R_P$, the ratio of tidal to perihelion radii), as well as the BH and WD masses.
Numerical simulations have established general criteria for robust ignition, namely
$\beta \gtrsim 3$ ($\gtrsim5$) 
for high (low) mass WDs \citep{Rosswog09, Tanikawa17, Kawana18, Anninos18, Anninos19}.
However it is well known that the scale height needed to
resolve the hydrodynamic flow perpendicular to the orbital plane scales nonlinearly with
tidal strength $\Delta \ell/R_{\mathrm{WD}} \propto \beta^{-3}$ \citep{Luminet86,Brassart08},
where $R_{\mathrm{WD}}$ is the WD radius, $\Delta \ell$ is the grid resolution off the plane, and we have
assumed a $\Gamma=5/3$ adiabatic gas law for the equation of state. For a
WD radius of $10^4$ km, a grid resolution of at least 10 km is thus required 
to model a $\beta=10$ interaction. This requirement has essentially been verified
with numerical models, though there is a practical limit as to
what can be done with computational resources in three dimensions, and it is
unclear how well this resolution scaling law carries over to reactive flows.
\citet{Tanikawa17} addressed this concern in part with 1D numerical calculations and provided 
general guidance for the minimum
resolution needed to achieve converged reactive thermonuclear solutions. They also
demonstrated that the appearance of shock networks and the sustainability of fuel cells
are both influenced strongly by the local gas density distribution and isotopic compositions at shock breakout,
something that 3D models cannot easily resolve.
Our previous models \citep{Anninos18, Anninos19} for example
did not fully cover, by roughly an order of magnitude, the spatial range
needed to determine precisely where the shocks first emerge or the role they play in either
initiating or facilitating thermonuclear burn. We were thus unable to comment conclusively on the
detonation mechanism (e.g., direct versus spontaneous initiation), the trigger
source (ballistic compression, pressure wave, shock), or on the hotspots
and induction gradients that develop on very fine scales.

Faced with a similar problem, the supernova community adopted a remap approach
for determining whether developing hotspots lead to successful detonation. The procedure involves extracting
partial solutions from multi-dimensional calculations meeting detection thresholds
(on density and temperature), then mapping those conditions onto a one-dimensional grid
to solve the reactive hydrodynamics equations at much higher resolution
\citep{Niemeyer97,Ropke07,Shen14}. The results from \citet{Seitenzahl09} in particular
have emerged as standard guidelines for the identification of
sustainable detonations in C/O white dwarf environments. That work
related critical hotspot length scales to composition (helium, oxygen, carbon mass fractions), 
thermal environments (background, peak temperatures), geometry, and functional distributions,
concluding that all of these effects conspire to produce a wide range of length scales,
from $10^{2}$ km to $10^{-2}$ km at densities $\sim10^6$ to $10^7$ g cm$^{-3}$.
\citet{Holcomb13} followed with a similar 1D analysis but for pure helium WD models
and simpler linear temperature profiles, finding somewhat less restrictive scales for ignition:
$10^2$, 1, and $10^{-2}$ km at respective densities $\sim10^6$, $10^7$ and $10^8$~g~cm$^{-3}$, a range
that encompasses our calculations.
It is not clear to what extent these results apply to tidal disruption events, but they offer reasonable guidelines
for the range of scales imposed by nuclear reactions.

Considering the sensitivity to model parameters and the underlying simplistic assumptions of 1D calculations,
we have undertaken in this work a high resolution computational study in two dimensions
that also reduces the parameter space by eliminating thermal environment,
geometry, and functional distribution uncertainties. Instead we allow the hydrodynamics
to develop those attributes from first principle evolutions, self-consistently with no 
ambiguous parameterization or remapping. As an added benefit, our parameter space is
reduced to just tidal strength and isotopic composition, where the latter also dictates the stellar mass.
We consider 0.6 (0.15) solar mass WDs with C/O (He) compositions, and fix
the black hole mass to $10^3 M_\odot$ for all calculations.

Section \ref{sec:methods} begins with a brief discussion of our numerical methods, physical models
(equation of state, reactive networks, initial data, etc.), and gridding strategy tuned to achieve the high
spatial resolution needed for resolving nucleosynthesis near the orbital plane.
Our results follow in Section \ref{sec:results}, and we conclude with a brief summary
in Section \ref{sec:conclusions}.

\section{Methods and Models}
\label{sec:methods}

All calculations are performed with the {\sc Cosmos++} code
\citep{Anninos05,Anninos12,Fragile14,Anninos17}, which solves
the equations of Newtonian or general relativistic hydrodynamics coupled with
thermonuclear reactions and energy generation on unstructured, moving and adaptively refined (AMR) meshes.  
Our previous work \citep{Anninos18,Anninos19} modeled ultra-close 3D encounters with general relativity
\added{
with perihelion separations as small as a few black hole gravitational radii $GM/c^2$. 
Instead this work solves the Newtonian equations (saving computational cost) since we are not looking to predict precise
trajectories where relativistic effects can deflect or precess stars off their Keplerian paths.
Our computational domain (in both space and time) is localized enough that precession effects can be neglected, and
the constrained 2D nature of the flow in any case prevents adequate resolution of orbital precession.
We do however account for the relativistic enhancement of conformal gravity and tidal forces with the Paczynsky-Wiita
\citep{Paczynsky80} potential which reproduces marginally stable and bound circular orbits around a Schwarzschild black hole:
$\Phi(r) = -GM/(r-R_S)$, where $R_S=2GM/c^2$ is the Schwarzschild radius.
Self-gravity is ignored due to the localized nature of these calculations where WD matter is initialized
well within the tidal radius of the BH ($\lesssim 0.5 R_T$, depending on the perihelion radius and
star parameters) so that tidal forces dominate at all times.
In particular the ratio of surface to tidal gravity (as defined for example
in \citet{Kochanek94}) ranges in our parameter studies between $10^{-4}$ and  $10^{-2}$.
In addition we have previously shown that self-gravity at these tidal distances has little effect on thermonuclear behavior,
either released energy or iron production,
even at distances twice the initial separation considered in this work \citep{Anninos18}.
}

Thermodynamics is treated with a Helmholtz equation of state,
accounting for electron degeneracy and relativistic and electron-positron contributions.
It is based on the Torch code \citep{Timmes99,Timmes00a} and
designed to work with arbitrary isotopic compositions and inline nuclear reaction networks.
The network is a 19-isotope $\alpha$-chain
and heavy-ion reaction model \citep{Weaver78,Timmes99,Timmes00b,Anninos18,Anninos19}
fully coupled with hydrodynamics, advection of reactive isotopics, photodissociation, and nuclear energy released.
\added{
Electron capture is ignored, but we do not expect those reactions to be important except at densities
roughly an order of magnitude greater than our models \citep{Langanke21}.
}

\subsection{Grid Parameters}
\label{subsec:gridParameters}

For comparison purposes we remind the reader our previous work had
developed a novel hybrid AMR plus mesh relaxation scheme to achieve
a dynamical range spanning approximately 4 decades in space (more if we were to include the
trajectory length scale) with a vertical resolution of $\lesssim10$ km, or equivalently
1/300th of a Schwarzschild radius. This resolution marginally satisfies the criteria
recommended by \citet{Tanikawa17} for converged nuclear solutions,
a criteria established through a series of
1D calculations run with more than an order of magnitude greater resolution than
our 3D models. The calculations presented here cover a spatial range comparable
to these 1D studies, spanning five decades of scale from the WD radius to sub-kilometer
zoning near the equatorial plane.
This is accomplished with static geometric zoning
(not adaptive regridding as in our previous calculations) which increases cell dimensions
in the vertical direction by the product $\Delta y_{i+1} = (1 + \delta) \Delta y_i$, where the constant factor ($\delta > 0$)
depends on the number of zones along the $y$-axis.
Most of the results presented in this report are derived
from grids of size 720$\times$720, but we use lower and occasionally higher cell counts to confirm convergence.
The actual physical resolution (cell width) depends of course on the number of cells, but also on
the interaction strength which relates the grid length to the tidal disruption time-scale as described below.

Stellar matter is injected onto the grid from the left side of the mesh at a scaled Keplerian
velocity $\epsilon_v v^x_K$, where $\epsilon_v$ is a multiplier used to synchronize peak compression to periapsis.
The vertical extent of the grid is fixed for all calculations at $L_Y = 1.2 R_{WD}$, where $R_{WD}$ is the star radius,
and along this direction we assign initial atomic compositions,
\added{densities, and energies in accordance with radial Mesa profiles \citep{Paxton11},
enforcing reflection boundary conditions at $y=0$,
and symmetry along the inflow boundary which allows the flow to relax in response to
black hole tidal forces.
}
\added{
Outside the star radius we impose a background density of $10^{-8}\rho_{\text{max}}$ and pressure $10^{-2} p_{\text{min}}$,
where $\rho_{\text{max}}$ and $p_{\text{min}}$ are the maximum density (at $y=0$)  and minimum pressure at the
star edge ($y=R_{WD}$). We maintain these floor values throughout the evolution.
}

The horizontal grid length $L_X$ and total simulation time $t_{sim}$ are correlated to the tidal disruption time-scale,
\begin{equation}
\tau_{TD} = \left(\frac{R_{P}^3}{G M_{BH}}\right)^{1/2}  ~,
\end{equation}
by defining $L_X = 4 v^x_K \tau_{TD}$ and $t_{sim} \gtrsim 4\tau_{TD}$.
The black hole is placed perpendicular to the two-dimensional plane in which stellar material moves,
offset at the specified perihelion radius $R_P$ but otherwise centered horizontally on the symmetry plane $y=0$.
Stellar matter thus enters the grid from the left and moves at roughly Keplerian speed across the grid,
first towards then away from the BH, while simultaneously undergoing tidal compression
in the vertical direction and a combination of stretching and compression in the horizontal direction from the black hole
located a distance $R_P$ off the trajectory plane. 

\added{
At or near periapsis, the effect of simplifying the flow to 2D eliminates a tidal force component that contributes to stretching
matter along the line of sight to the black hole. Elimination of this component implies that the peak densities might be slightly
greater in 2D compared to 3D. However because of the relatively short spatial scales of these calculations together with the fast orbital velocity,
the time to transit through the transverse compression phase at periapsis is comparable to the tidal disruption timescale,
so we don't expect this to impact our results or conclusions much. We verified this by comparing
on-axis densities with our previous 3D calculations, confirming they are roughly equivalent.
More broadly we have verified that this 2D simplification
reproduces the dynamical behavior and internal structure (e.g., density, temperature, nozzle compression, 
shock generation, and pressure relaxation) found in 3D tidal disruption models.
}

\subsection{Model Parameters}
\label{subsec:modelParameters}

Apart from black hole spin, the models and interaction parameters are otherwise
similar to those adopted in \citet{Anninos19}: 
a $0.6 M_\odot$ (C/O) and a $0.15 M_\odot$ (He) WD approaching a $10^3 M_\odot$ BH.
The 0.15 $M_\odot$ model is composed of essentially a uniform distribution
of 99\% mass fraction $^4$He with trace amounts of other species.
The 0.6 $M_\odot$ model is composed of a central region with roughly a homogeneous
mixture of 1/3 $^{12}$C and 2/3 $^{16}$O and trace amounts of heavier nuclei, surrounded in turn by layers
of carbon-rich, helium-rich, and hydrogen-rich material, ordered from the inner core to the outer surface.
\added{
Trace materials include nitrogen, magnesium, neon, and silicon distributed according to Mesa-generated
radial profiles with mass fractions of order $10^{-3}$ or less.
}

We choose the remaining free parameter, the perihelion radius, to produce
the range of tidal strengths ($\beta$) shown in Table \ref{tab:runs}, where
$\beta$ is the ratio of tidal to perihelion radii $\beta=R_T/R_P$, with tidal radius
\begin{eqnarray}
R_\mathrm{T} &\approx& 1.2\times10^{5} \left(\frac{R_\mathrm{WD}}{10^9 \mathrm{cm}}\right)
                       \left(\frac{M_\mathrm{BH}}{10^3 M_\odot}\right)^{1/3}
                       \left(\frac{M_\mathrm{WD}}{0.6 M_\odot}\right)^{-1/3}	~\mathrm{km} ~,	\\
             &\approx& 40 \left(\frac{R_\mathrm{WD}}{10^9 \mathrm{cm}}\right)
                       \left(\frac{M_\mathrm{BH}}{10^3 M_\odot}\right)^{-2/3}
                       \left(\frac{M_\mathrm{WD}}{0.6 M_\odot}\right)^{-1/3}	~R_S ~,
\end{eqnarray}
for a black hole mass $M_\mathrm{BH}$, stellar radius $R_\mathrm{WD}$, stellar mass $M_\mathrm{WD}$,
and Schwarzschild radius
\begin{equation}
R_S = \frac{2GM_\mathrm{BH}}{c^2} \approx 3\times10^3 \left(\frac{M_\mathrm{BH}}{10^3 M_\odot}\right) ~\text{km}  ~.
\end{equation}
These encounter scenarios are violent enough to trigger robust nucleosynthesis 
(exceeding $\beta > 3$ for CO WDs, and $\beta > 5$ for He WDs) without being
overly restrictive on grid resolution, which scales as $\Delta y \propto \beta^{-3}$.
In addition, the upper ends for both C/O and He WDs are approximately
at the capture limit \citep{Kawana18}
\begin{equation}
\beta_{max} \lesssim 10 
                  \left(\frac{R_\mathrm{WD}}{10^9 \mathrm{cm}}\right)
                  \left(\frac{M_\mathrm{BH}}{10^3 M_\odot}\right)^{-2/3}
                  \left(\frac{M_\mathrm{WD}}{0.6 M_\odot}\right)^{-1/3}	~,
\end{equation}
beyond which much of the debris is expected to be bound to the BH,
so they are reasonable maxima for our studies.

Table \ref{tab:runs} lists our ensemble model parameters, including
star masses ($M_{WD}$ in solar mass units), star radii ($R_{\mathrm{WD}}$ in units of $R_S$), 
perihelion radii ($R_P$ in units of $R_S$), and interaction strengths $\beta$.
The run labels denote physical parameters: The first upper case letter `M' signifies the WD mass
(M2 $\equiv0.15~M_\odot$, M6 $\equiv0.6~M_\odot$), the second number following upper case letter `R' is
the perihelion radius in units of $R_S$.

\begin{deluxetable}{lcccccc}
\tablecaption{Run Parameters \label{tab:runs}}
\tablewidth{0pt}
\tablehead{
\colhead{Run}             & 
\colhead{$M_\mathrm{WD}$} & 
\colhead{$R_\mathrm{WD}$} & 
\colhead{$R_\mathrm{P}$}  & 
\colhead{$\beta$}     \\
                          & 
($M_\odot$)               & 
($R_S$)                   & 
($R_S$)                   & 
}
\startdata
\hline
M2R4      &  0.15 & 5.4 &  4 &  23  \\
M2R8      &  0.15 & 5.4 &  8 &  12  \\
M2R12     &  0.15 & 5.4 & 12 &   8  \\
M6R4      &  0.6  & 2.9 &  4 &   9  \\
M6R8      &  0.6  & 2.9 &  8 &   4 
\enddata
\end{deluxetable}

\section{Results}
\label{sec:results}

\begin{figure}
\hspace{-1in}\includegraphics[width=1.3\textwidth]{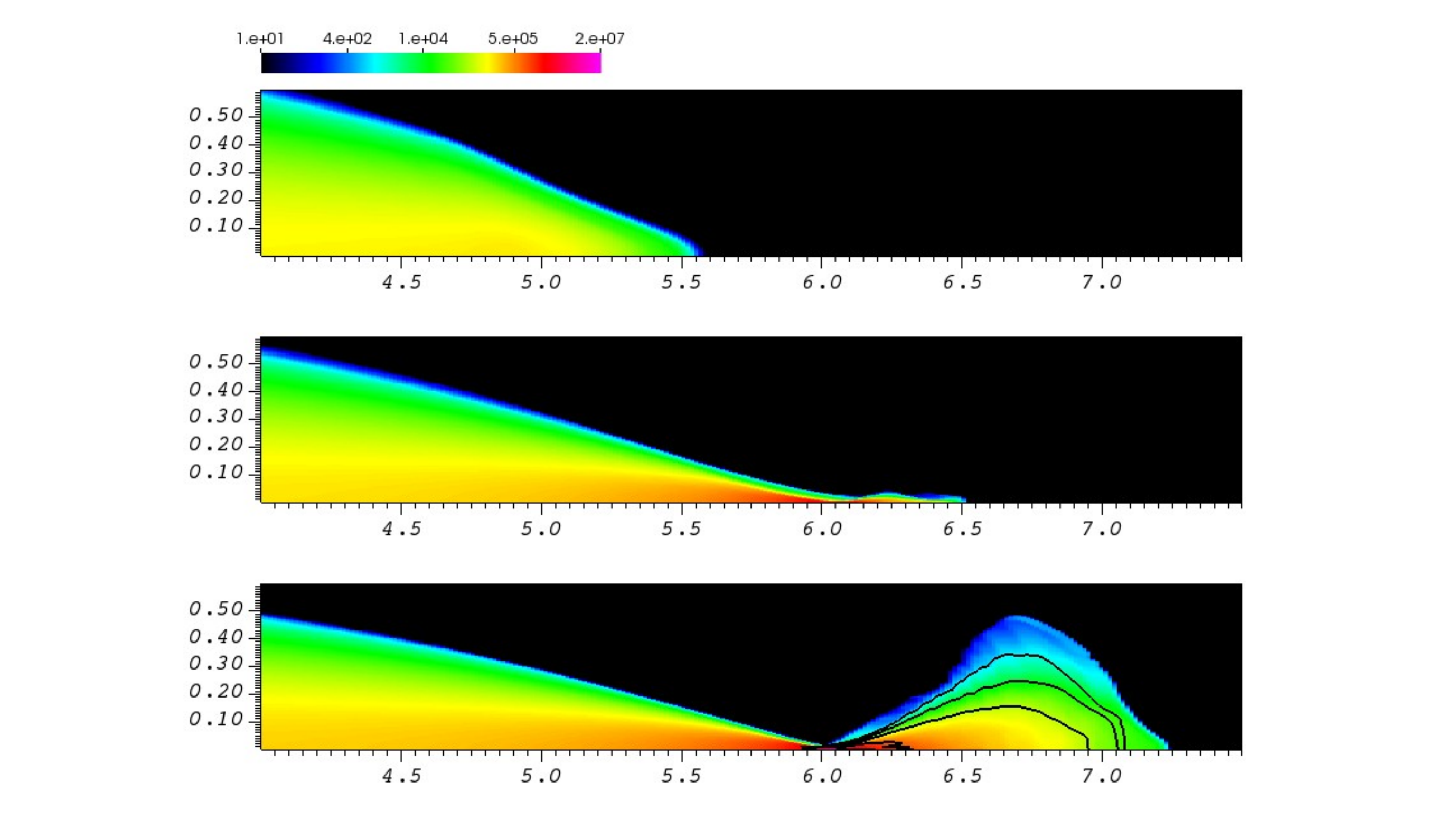}
\caption{
Logarithm of the mass density in cgs units for case M2R8 at the following times: 
1.25 (top), 1.36 (middle) and 1.45 (bottom) seconds following the entry of matter
through the left grid boundary. The density color bar is truncated to six orders
of magnitude to enhance the structure of the fluid stream (the actual
minimum density in the calculations is roughly four orders of magnitude smaller).
The gas propagates to the right at roughly Keplerian velocity
while responding to tidal forces from a $10^3 M_\odot$ BH located
off the 2D plane by a distance of $R_P = 8 R_S$, but otherwise centered
on the grid at $y=0$. The top image is an early time as the WD begins to compress,
the middle image is just when the WD reaches maximum compression and shocks off
the orbital plane (observed at the front of the flow), 
and the bottom image shows the subsequent bounce driving debris off the orbital plane.
Also shown in the bottom image are contours of the nickel density at logarithmic
intervals $5\times10^2$, $5\times10^3$, $5\times10^4$, and $5\times10^5$ g/cm$^3$.
One unit length scale corresponds to $10^4$ km.
}
\label{fig:images_den_M2R8}
\end{figure}

\begin{figure}
\includegraphics[width=1.0\textwidth]{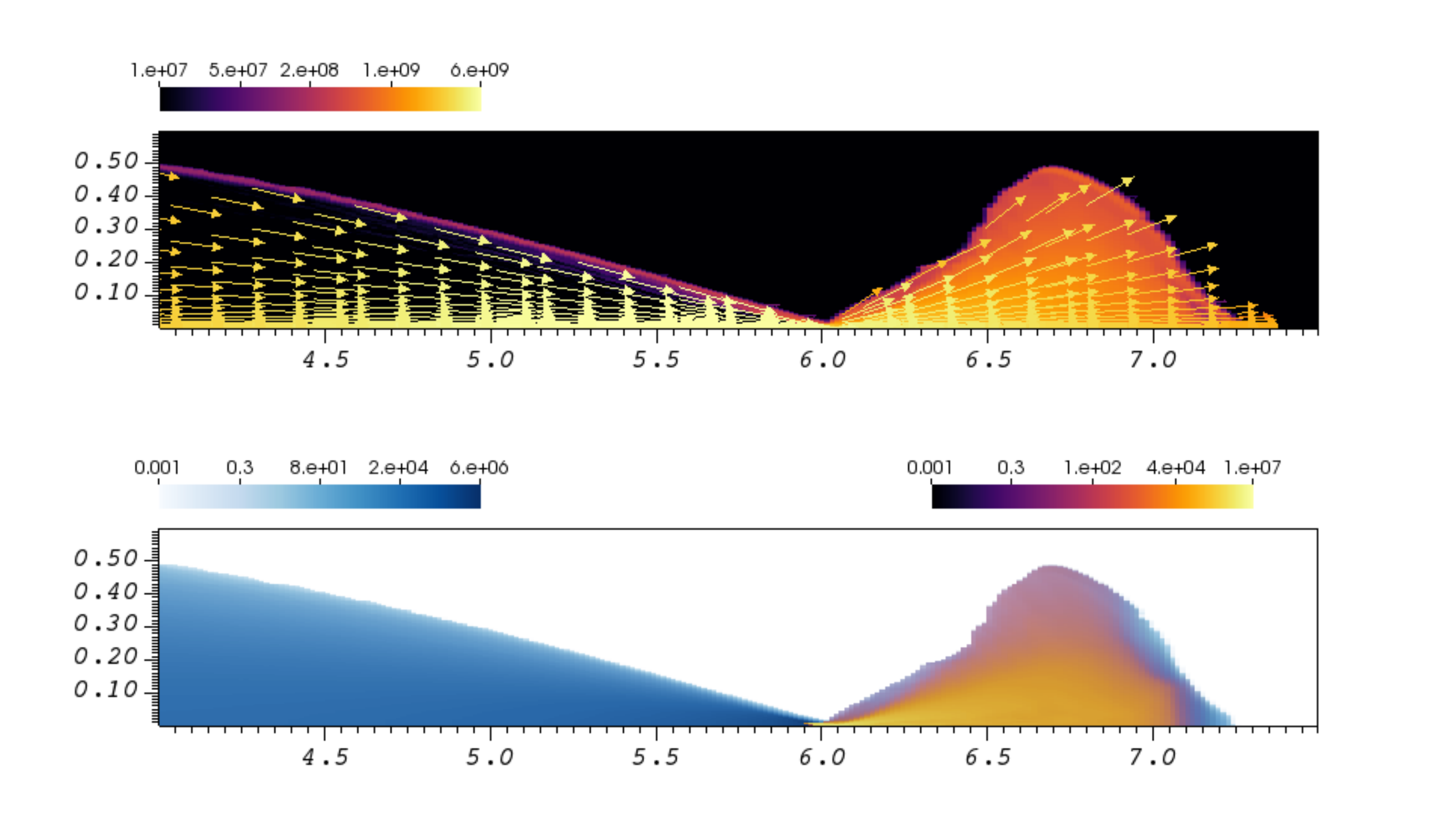}
\caption{
Logarithm of the gas temperature (top image) and helium and nickel densities (bottom image)
for the same case and final time as Figure \ref{fig:images_den_M2R8}. The temperature is plotted
in degrees Kelvin, the densities are cgs represented by ramped opacity color maps (blue for He, inferno for Ni).
The top image additionally shows the velocity flow field (arrows) scaled to
the magnitude, which is nearly uniform over this domain $\sim9\times10^9$ cm/s.
One unit length scale corresponds to $10^4$ km.
}
\label{fig:images_tem_M2R8}
\end{figure}

Figure \ref{fig:images_den_M2R8} shows time sequences of the logarithm of the gas density for a representative
(intermediate perihelion) case M2R8. The top image corresponds to an early time,
1.25 seconds after material enters the grid from the left. Material travels along
the positive x-direction at roughly Keplerian velocity, experiencing tidal stretching and compression
forces from the black hole centered horizontally at $y=0$, but displaced
off the 2D plane by a distance equal to the perihelion radius ($R_P = 8 R_S$ in this example).
Over time, matter converges onto the orbital plane at $y=0$ and eventually reaches maximum compression
and detonates, as observed at the head of the flow in the second (middle) image at 1.36 seconds.
The last (bottom) image corresponds to a time of 1.45 seconds, after material 
rebounds off the orbital plane and commences to disperse, less than a
tenth of a second following peak compression.
Nucleosynthesis begins in earnest
just before peak compression at or near the orbital plane, and very quickly builds burn product
chains terminating with $^{56}$Ni. The bottom image in Figure \ref{fig:images_den_M2R8}
additionally plots logarithmic contours of the $^{56}$Ni density showing the prominent production sites.
Figure \ref{fig:images_tem_M2R8} shows the corresponding temperature and velocity field (top image), and helium
and nickel densities (bottom image) at the same final time as Figure \ref{fig:images_den_M2R8}.

These images are typical of all the case studies we have run. The main differences are the
horizontal length scale over which compression takes place, the shock break-out positions, and the
peak densities and temperatures, all of which depend on the perihelion radius
and play strong roles in the production and distribution of iron group elements.

Although this behavior is articulated here in two dimensions it is representative of what
we find in more realistic 3D encounters. Heavy element synthesis, for example,
initiates abruptly in 3D geometries in (or near) the orbital plane the moment
stellar matter compresses to ignition densities and temperatures.
This occurs primarily along a radially aligned filamentary-like structure
which forms at the cusp of peak compression, and whose two-dimensional cross-section appears
much like Figures \ref{fig:images_den_M2R8} and \ref{fig:images_tem_M2R8}, resembling nozzle flow.
Detonation shocks develop in both 2D and 3D near the
orbital plane and propagate upwards and outwards, leaving nuclear ash in their wake.
Similarities and differences between 2D and 3D will be emphasized in subsequent sections as
we discuss the 2D results in greater detail.

\subsection{Helium WD Results}
\label{subsec:hewd}

\subsubsection{Peak Diagnostics}
\label{subsubsec:bulk_he}

\replaced
{
Table \ref{tab:peakdiag} summarizes key diagnostics from the low mass He WD
calculations, including the maximum density ($\rho_\mathrm{max}$) and
temperature ($T_\mathrm{max}$) on the grid, as well
as the total nuclear energy generated ($e_{nuc}$) and the
maximum mass of iron ($M_\mathrm{Fe, max}$) and calcium ($M_\mathrm{Ca, max}$) group elements.
We define iron (calcium) group as those elements with atomic numbers in the range 52-60 (40-51).
For the 19-isotope network used in our calculations, these group elements consist simply of
(Fe, Ni) and (Ca, Ti, Cr), respectively.
}
{
Table \ref{tab:peakdiag} summarizes key diagnostics from the low mass He WD
calculations, including the maximum density ($\rho_\mathrm{max}$) and 
temperature ($T_\mathrm{max}$) on the grid, as well
as the total nuclear energy generated ($e_{nuc}$) and the
maximum mass fractions of iron/nickel elements ($\mathrm{Fe, max}$) and calcium ($\mathrm{Ca, max}$).
}

Our current model parameters differ somewhat
from \citet{Anninos18}, but one can nonetheless draw similarities between pairs (M2R4, B3M2R06) for the strongest
interaction strengths, pairs (M2R8, B3M2R09) for the intermediate strength interactions,
and an average between B3M2R09 and B3M2R20 for the most distant encounter M2R12, where the
runs labeled ``M'' (``B'') refer to our new (old) calculations.
The maximum densities and temperatures in \citet{Anninos18} are averaged
over the densest 10\% of WD matter, whereas here they are peak zonal quantities. 
Hence we expect the values in Table \ref{tab:peakdiag} to be slightly greater, as they indeed are.
Despite this difference in definition, they (along with the other diagnostics) 
agree for the most part to within a factor of a few, providing confidence
that these 2D models are excellent proxies for 3D geometries while accommodating high resolution studies.

\begin{deluxetable}{lcccccc}
\tablecaption{Peak Diagnostics \label{tab:peakdiag}}
\tablewidth{0pt}
\tablehead{
\colhead{Run}                     & 
\colhead{$\beta$}                 & 
\colhead{$\rho_\mathrm{max}$}     & 
\colhead{$T_\mathrm{max}$}        &
\colhead{$e_\mathrm{nuc}$}        & 
\colhead{$\mathrm{Fe, max}$}   & 
\colhead{$\mathrm{Ca, max}$}   \\
                                  & 
                                  & 
(g cm$^{-3}$)                     & 
(K)                               &
(erg)                             & 
(mass fraction)                   & 
(mass fraction)
}
\startdata
M2R4   & 23 & $9.6\times10^7$ & $1.0\times10^{10}$ & $1.8\times10^{50}$ & $2.9\times10^{-1}$ & $2.5\times10^{-4}$ \\
M2R8   & 12 & $5.3\times10^7$ & $5.2\times10^{ 9}$ & $6.2\times10^{49}$ & $1.1\times10^{-1}$ & $2.0\times10^{-4}$ \\
M2R12  &  8 & $1.3\times10^7$ & $4.1\times10^{ 9}$ & $4.1\times10^{49}$ & $6.8\times10^{-2}$ & $4.6\times10^{-4}$ \\
M6R4   &  9 & $3.9\times10^8$ & $1.3\times10^{10}$ & $3.6\times10^{50}$ & $3.9\times10^{-1}$ & $1.3\times10^{-3}$ \\
M6R8   &  4 & $1.5\times10^8$ & $6.9\times10^{ 9}$ & $1.5\times10^{50}$ & $1.7\times10^{-1}$ & $1.2\times10^{-3}$ \\
\enddata
\end{deluxetable}

\subsubsection{Initiation}
\label{subsubsec:initiation_he}

\begin{figure}
\includegraphics[width=0.9\textwidth]{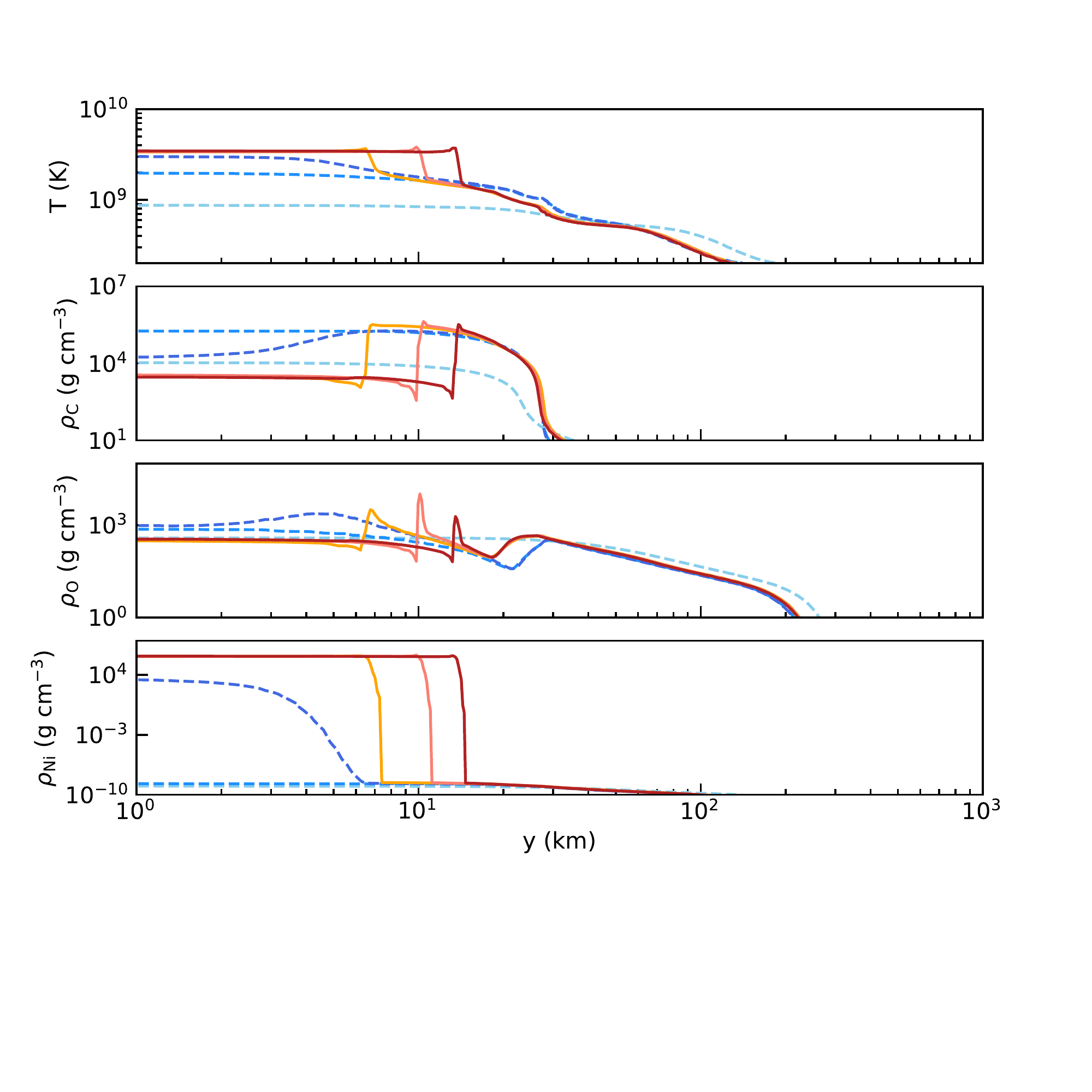}
\caption{
Line profiles showing the temporal evolution of the temperature (top), carbon density, oxygen density, and nickel (bottom)
along the vertical direction ($y$-axis) for case M2R8.
Blue dashed (orange solid) lines correspond to solutions at different times before (after) the detonation producing iron elements.
Transitions from light to dark tones represent sequences from early to late times.
}
\label{fig:1D_rho_M2R8}
\end{figure}

The evolving thermodynamic state of the tidal stream
is represented in Figure \ref{fig:1D_rho_M2R8} as a chronological sequence of one-dimensional line
profiles (before and after detonation) of
temperature, carbon density, oxygen density, and nickel along the vertical $y$-axis perpendicular to the orbital plane.
Results are shown for the same case M2R8 as the 2D color plates. Each individual colored curve
corresponds to a different time before and after the detonation producing nickel.
The pre- (post-) detonation solutions are plotted as a family of dashed blue (solid orange ) lines,
and transitions from light to dark tones represent sequences from early to late times.
\added{
To get a sense of the scales involved we note that the blue (orange) lines, ordered
light to dark, correspond to times 1.338, 1.344, 1.3444 (1.3456, 1.346, 1.3463) seconds.
}
The horizontal $x$-positions at which these line profiles are extracted evolve in time
with the maximum hydrodynamic pressure on the grid. So they are not exactly
Lagrangian tracers but do move along the positive $x$ direction tracking the location of peak compression.
Furthermore, we note that the last dashed curve and the first of the solid lines are extracted from consecutive time dumps,
so the transition to a detonation occurs over significantly shorter timescales than the frequency at which we store data
to disk ($10^{-4}~\tau_{TD} \sim\text{few}\times 10^{-4}$ seconds, varying slightly model to model).

The pre-detonation phase begins with homologous collapse which drives adiabatic compression until the
density and temperature increase to where carbon is made through the triple alpha reaction. The burning
of helium to carbon occurs mildly without a shock or detonation. Energy released in the process enhances adiabatic heating,
further slowing and stagnating the flow off the orbital plane. Initially even nickel is produced mildly without
much disruption to the hydrodynamic flow. However, as the gas continues to heat and
nickel production increases, conditions (density and temperature) evolve to trigger a detonation wave strong
enough to burn carbon, oxygen, and transient products quickly to nickel and iron elements.

\begin{figure}
\includegraphics[width=0.6\textwidth]{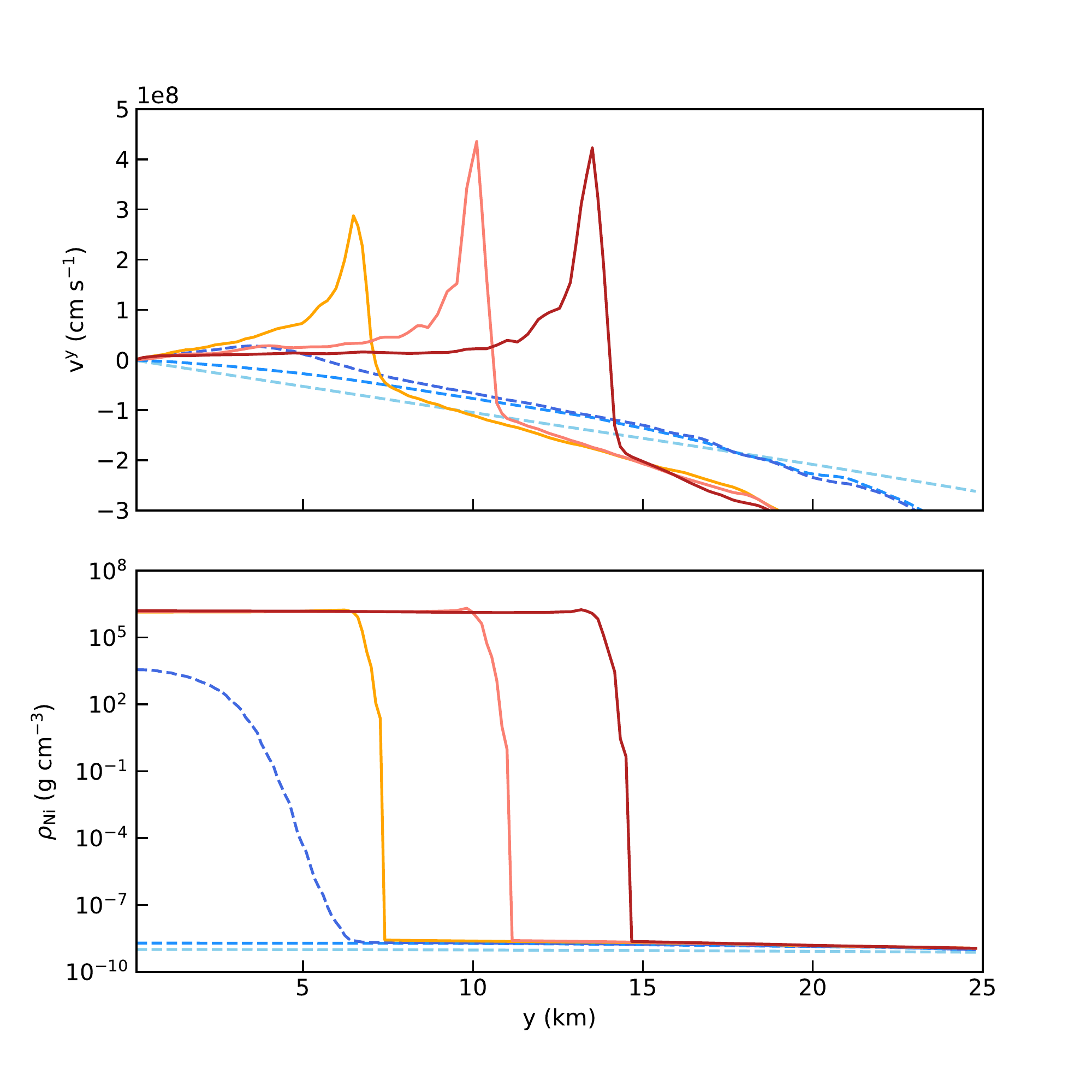}
\caption{
Vertical velocity component (top) and nickel density for case M2R8 shown at the same times as Figure \ref{fig:1D_rho_M2R8}. 
}
\label{fig:1D_velrho_M2R8}
\end{figure}

As nickel production increases, the released energy
begins at some point to drive an outflow from the central plane that eventually evolves to a shock and detonation wave.
This is demonstrated by Figure \ref{fig:1D_velrho_M2R8} where we plot the vertical component of velocity
in the top plate and nickel density in the bottom, zoomed close to the central plane. The first two (light blue) 
dashed curves represent early carbon burn and are smooth,
displaying a transition from homologous collapse to enhanced adiabatic compression. The remaining later time
curves show the outflow (dark blue) and velocity steepening (orange) both correlate strongly with the onset of nickel production.
The transition to detonation happens for this particular case
at $\sim 2\times10^6$ g cm$^{-3}$ and $\sim2.5\times10^9$ degrees Kelvin in a region extending 7 km off the orbital plane.
The size of this developing hotspot ($\sim$14 km above and below the plane), whether measured as the extent of the early carbon front
or the early site of detonation, is consistent with the critical length scale constraints established by \citet{Holcomb13} at this density.

From Figure \ref{fig:1D_rho_M2R8} we can see the early carbon build-up phase proceeds slowly and smoothly 
(and profiles of the vertical velocity in Figure \ref{fig:1D_velrho_M2R8} bear that out), but a lot of it is made over time.
The steady state composition ratio of carbon to helium in these low mass stellar models does not exceed $10^{-4}$ throughout the star.
However, just prior to detonation carbon grows to represent roughly 10\% of the compressed high density material as it preconditions the
core for ignition, with pockets of higher mass products (e.g., calcium) growing to similar levels, but over smaller localized domains.
Once detonation triggers much of the carbon (and oxygen) converts to iron as the shock wave overtakes the slower moving carbon reactive front.
The reaction front in this example
eventually propagates out past 100 km above the orbital plane, affecting a large fraction of the tidal stream.

\begin{figure}
\includegraphics[width=0.6\textwidth]{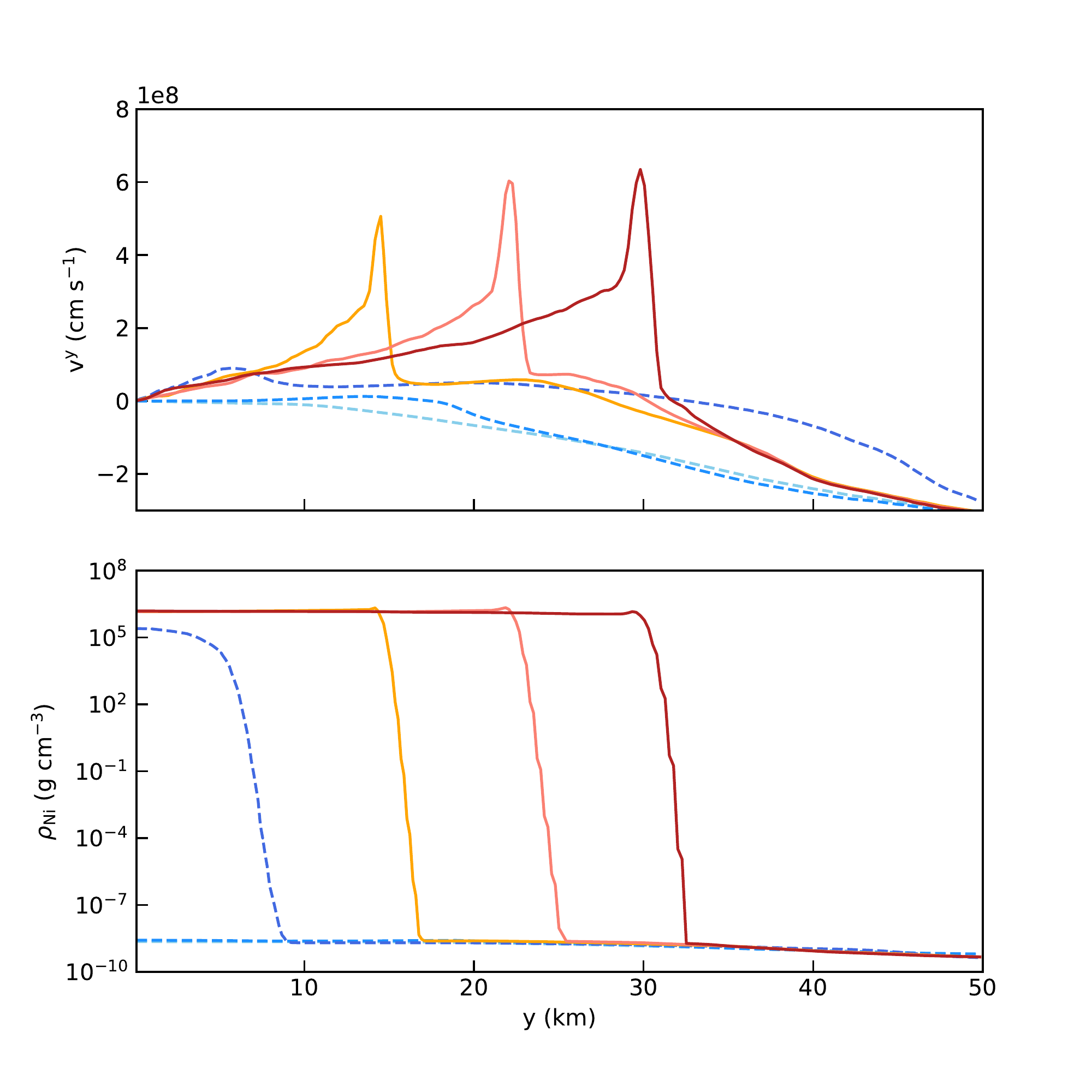}
\caption{
As Figure \ref{fig:1D_velrho_M2R8} but for case M2R12.
}
\label{fig:1D_velrho_M2R12}
\end{figure}

The most distant He WD encounter scenario M2R12 behaves similar to M2R8, but for following a trend whereby
detonation occurs progressively further off the orbital plane with a clearer and more easily identifiable
role of hydrodynamic outflows in the initiation process. The temperature and density profiles are similar
to Figure \ref{fig:1D_rho_M2R8} except the detonation front first appears just outside of 10 km off the plane.
Like M2R8, the lead-up to detonation occurs by preconditioning the fuel with mild helium burn to carbon,
increasing the gas density and temperature to ignition conditions $\sim 2\times10^6$ g cm$^{-3}$ and $\sim 2.5\times10^9$ K.
The early phase of nickel production occurs mildly as well, until enough of it is created to support
explosive burn and a detonation shock. Figure \ref{fig:1D_velrho_M2R12} shows much of the same behavior as
Figure \ref{fig:1D_velrho_M2R8}, but also finds a significantly broader and faster outflow is generated
just as nickel production begins, presumably due to the delayed onset of detonation (compared to M2R8)
which allows more time for the outflow to develop and strengthen by accretion. This outflow eventually
steepens to a shock as nickel production accelerates and more energy is liberated at increasingly faster rates.

\begin{figure}
\includegraphics[width=0.8\textwidth]{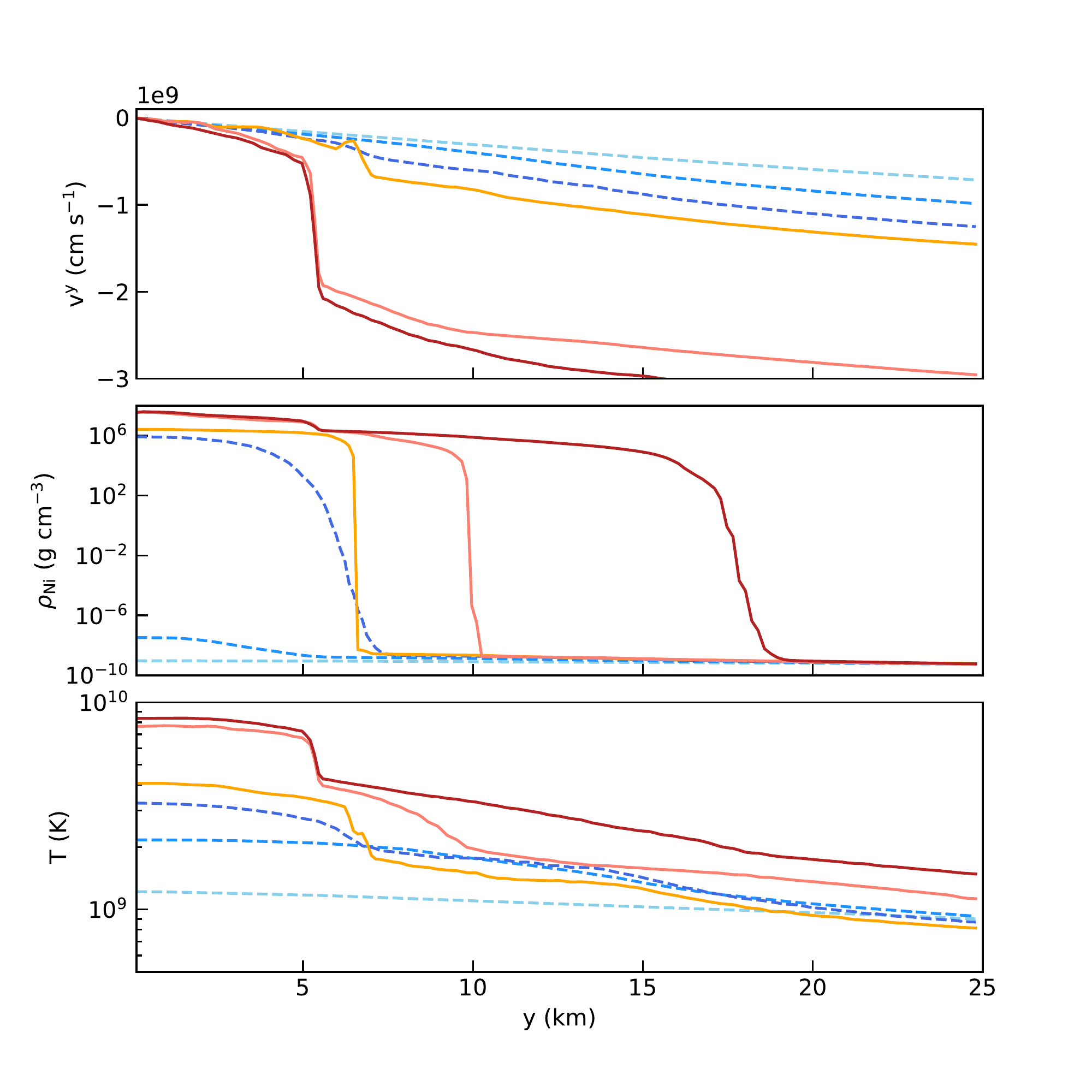}
\caption{
Vertical velocity component (top), nickel density (middle), and temperature (bottom) 
for the strongest tidal interaction case M2R4.
}
\label{fig:1D_velrho_M2R4}
\end{figure}

The nearest encounter case M2R4, represented by Figure \ref{fig:1D_velrho_M2R4}, is an extreme example where detonation appears in the early phases
to evolve like M2R8, where carbon, oxygen, and even nickel are created mildly through a combination of
homologous collapse and adiabatic compression heating from nuclear energy released. However in this case
we don't observe an outflow, even from saturated nickel production. A shock does form at the
initiation site where the nickel reaction front steepens, but the reaction front and shock separate
as the shock remains more or less stationary near the activation site roughly 5 km from the orbital plane.
By the time the central region detonates, as much as 10\% of the total mass contained in the 5 kms
is made up of iron group elements.
Added pressure from energy released behind the front is not sufficient to overcome the stronger tidal force
(and faster transverse velocities), making it difficult
for the hydrodynamic shock to keep pace with the reactive front even as more gas falls through the shock
at increasingly supersonic speeds.

\begin{figure}
\includegraphics[width=0.7\textwidth]{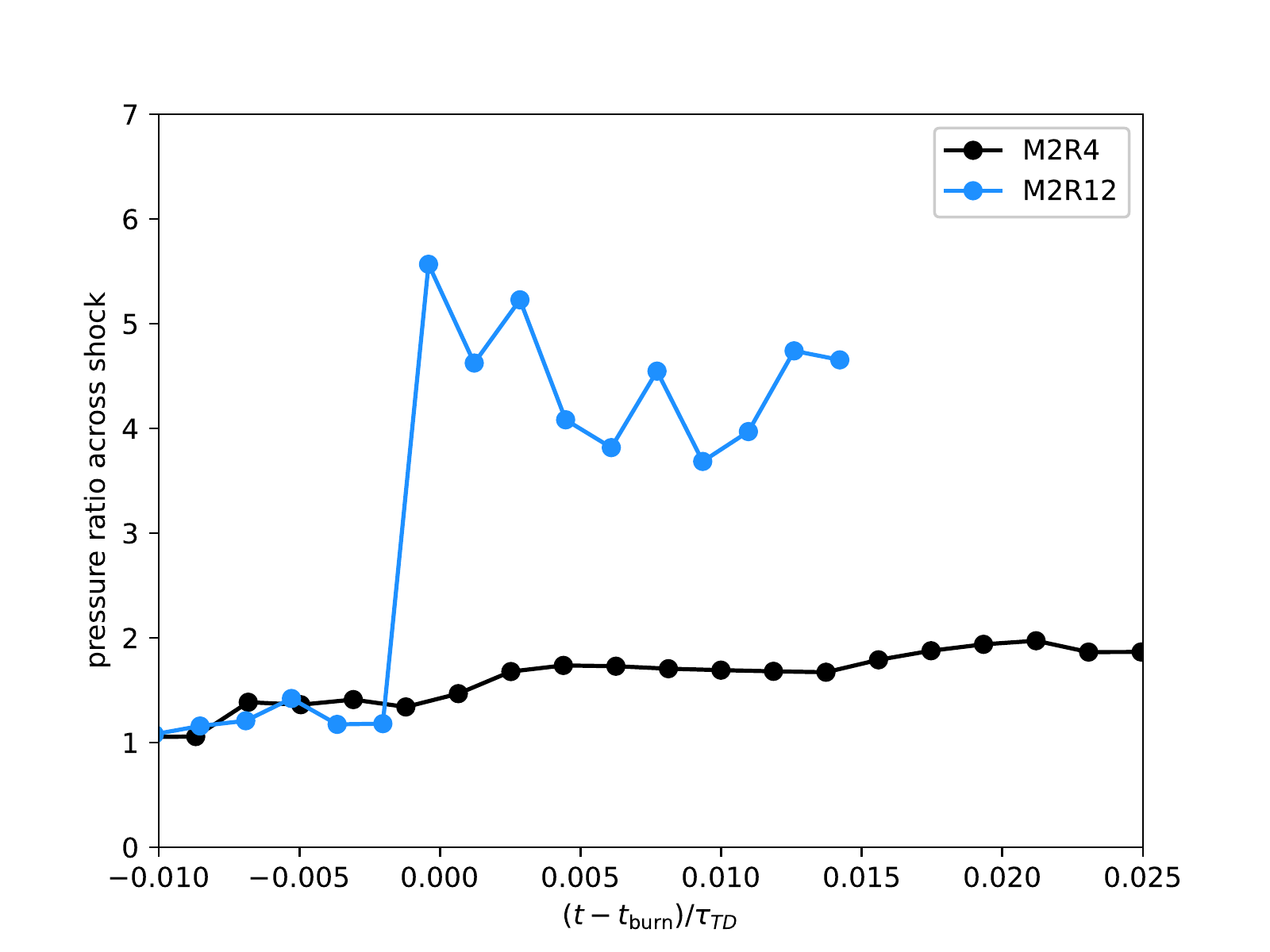}
\caption{
Ratio of total pressures across the hydrodynamic shock for two of the He WD models. The numerator (denominator) of the ratio 
corresponds to the downstream shocked (upstream)  material closer to the midplane. Both thermal and ram pressure contributions are included, 
although thermal pressure dominates for the shocked material and ram pressure dominates for the upstream material. 
The pressures have been averaged over the plotted time intervals ($\approx 10^{-3} \, \tau_{TD}$). 
}
\label{fig:pressure_ratio}
\end{figure}

Figure~\ref{fig:pressure_ratio} examines this in more detail. The figure shows, as a function of time, the downstream to upstream pressure ratio across the shock, for the M2R4 and M2R12 models. The downstream (numerator) of this ratio corresponds to the region closer to the midplane, and the shock
location is identified by the jump in thermal pressure. The time coordinate in this plot has been adjusted so that
zero corresponds approximately to the peak nickel production rate.
Both thermal and ram pressure contributions are included, although thermal pressure dominates 
for the shocked material and ram pressure dominates for the upstream material. 
We see the thermal pressure of the nuclear ash in M2R12 eventually exceeds the ram pressure of the infalling material by factors of 4 or more, causing the shock to move out along with the Ni-burning reaction front.  However, for M2R4, the pressure ratio across the shock remains closer to unity. 
We attribute this to two-dimensional BH proximity effects (stronger vertical tidal pull and faster transverse velocities) that
conspire to prevent the shock from propagating as high off the orbital plane as the more distant encounter case.
As a result, for the M2R4 case, the shock slows the incoming gas from supersonic to subsonic vertically
directed velocities, but does not generate outflow.
In the meantime, the reaction front in the M2R4 case decouples from the shock and moves upward and outward, partly by nuclear energy produced
but also by accretion which increases the density and temperature of the central region. In fact, the edge of the nickel
density curves in Figure \ref{fig:1D_velrho_M2R4} line up with and track the position at which the gas achieves a density
and temperature of $\sim10^6$ g cm$^{-3}$ and $\sim 2.5\times10^9$ K.
Eventually, the reaction front in this case stalls less than 50 km from the orbital plane due to blow-out after propagating into
steep density/temperature gradients where burn cannot be sustained, a distance that is significantly less than
the other model by a factor of two.

\begin{figure}
\includegraphics[width=0.7\textwidth]{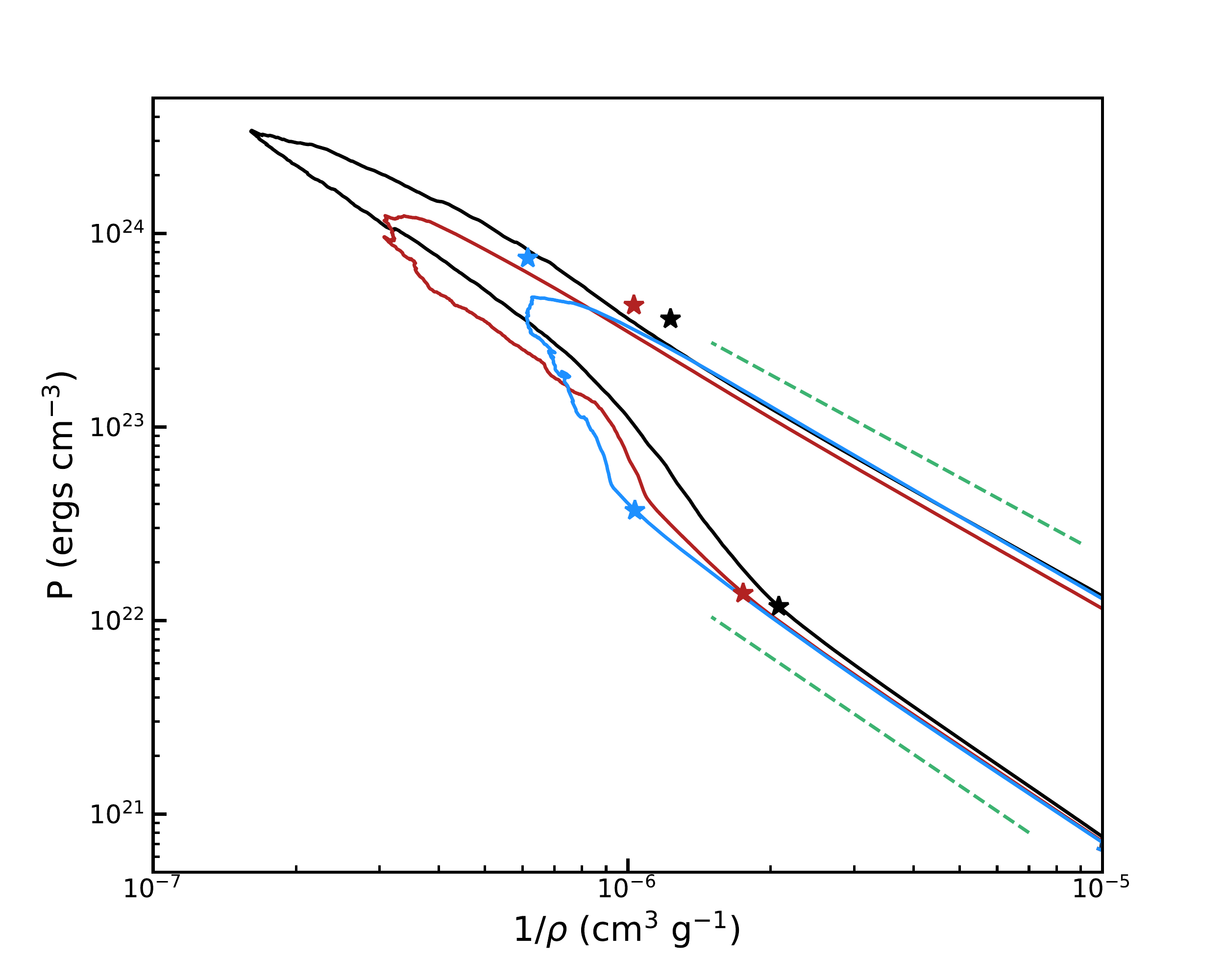}
\caption{
Average Hugoniot tracks of the pressure as a function of specific volume
experienced by a collection of Lagrangian tracer particles through the
compression, explosion, and decompression phases.
The different colored solid curves represent results from the three different
cases: M2R4 (black), M2R8 (red), and M2R12 (blue). The star markers correspond to jump conditions for
Chapman-Jouguet detonations, under the assumptions and approximations stated in the main text. 
The lower and upper green dashed lines are the gamma-law behaviors corresponding to
gas ($\gamma=5/3$) and radiation ($\gamma=4/3$) pressure dominated regions respectively.
}
\label{fig:phase_prhoi}
\end{figure}

Finally we plot in Figure \ref{fig:phase_prhoi} the average pressure-specific volume (Hugoniot) phase histories experienced by
an ensemble of Lagrangian tracer particles from all three models as they collapse towards the orbital plane, cross the
detonation front, then decompress at late times. The tracks begin at
low density and pressure (in the bottom right of the plot), then move upwards and
to the left as stellar material collapses adiabatically to higher density and pressure.
Nuclear burn occurs where the Hugoniot deviates from the $\gamma \sim 5/3$ adiabatic compression rate
(lower green dashed line) at a density of about $10^6$ g cm$^{-3}$. The pressure and density continue to build
from a combination of tidal compression and nuclear energy generation until a critical point
is reached where the internal pressure build-up overcomes gravitational forces. The subsequent
decompression phase is driven primarily by radiation pressure
at high post-detonation temperatures, and leaves a clear signature
on the Hugoniot as the gas relaxes at a shallower $\gamma \sim 4/3$ rate (top green dashed line).

In Figure~\ref{fig:phase_prhoi} we have also included star markers to indicate jump conditions 
for an idealized Chapman-Jouguet detonation corresponding to the three simulations. These conditions are
derived assuming a pure Helium-4 composition for the unburnt material that converts 60\% of its mass to nickel 
in the downstream material, while the nuclear energy released is completely thermalized. 
This burning efficiency is in good agreement with the actual computed nickel mass fractions seen in the 
burnt material within a time of 0.0025 $\tau_{TD}$ following the onset of nuclear burning. 
For the gas equation of state, we account for radiation pressure, electron pressure including the effects of degeneracy, and ideal gas pressure for the ions. We see that these idealized jump conditions 
predict consistently higher ash pressures compared to their corresponding simulated Hugoniot tracks. By this criterion, the simulated burn fronts can be classified as relatively ``weak detonations.''
We infer that competing effects from vertical tidal compression (driving the detonation) and transverse motion
(affecting pressure and kinetic relaxation of the ash phase) are responsible for the overall weakening of the burn 
front compared to an equivalent steady-state, one-dimensional detonation with the same upstream densities and temperatures.

\subsubsection{Convergence Studies}
\label{subsubsec:convergence_he}

Considering the small scales over which detonations
are triggered, we evaluate in this section the sensitivity of numerical
calculations to spatial resolution. This is of particular concern for assessing the
accuracy and reliability of three dimensional calculations.

Table \ref{tab:results_conv} summarizes results from
a series of convergence studies performed along both horizontal and vertical directions.
The usual diagnostics (density, temperature, energy, burn products) are tabulated as a function
of the number of zones ($N_x$, $N_y$) and minimum 
spatial resolution ($\Delta x$, $\Delta y$) in each spatial direction.
We find virtually no difference in the total energy released nor iron group mass fractions
by varying the horizontal resolution $\Delta x$. There is some sensitivity in the
peak calcium mass, but it is less than linear.

The resolution study carried out in the vertical direction is more interesting. We find
little variation in either the total energy released or the total iron products:
less than 20\% for both diagnostics at resolutions between 0.2 to 100 km,
but roughly three times that for calcium abundances.
At resolutions worse than 100 km, errors increase significantly. 
At resolutions better than 10 km, the more robust attributes are converged
to a few percent, while transient products remain uncertain to about 50\%.
The 10 km recommendation proposed by \citet{Tanikawa17} is therefore a reasonable
target for numerical models, except perhaps when precise knowledge of transient
yields is required.

\begin{deluxetable}{lccccccccc}
\tablecaption{Convergence Studies \label{tab:results_conv}}
\tablewidth{0pt}
\tablehead{
\colhead{Run}                     & 
\colhead{$N_x$}                   & 
\colhead{$N_y$}                   & 
\colhead{$\Delta x$}              & 
\colhead{$\Delta y_\mathrm{min}$} & 
\colhead{$\rho_\mathrm{max}$}     & 
\colhead{$T_\mathrm{max}$}        &
\colhead{$e_\mathrm{nuc}$}        & 
\colhead{$\mathrm{Fe, max}$}   & 
\colhead{$\mathrm{Ca, max}$}   \\
                                  & 
                                  & 
                                  & 
(km)                              & 
(km)                              & 
(g cm$^{-3}$)                     & 
(K)                               &
(erg)                             & 
(mass fraction)                   & 
(mass fraction)
}
\startdata
M2R4  & 720 & 720 &  77 & 0.2 & $9.7\times10^7$ & $1.0\times10^{10}$ & $1.8\times10^{50}$ & $2.9\times10^{-1}$ & $2.5\times10^{-4}$ \\
      & 360 & 720 & 154 & 0.2 & $7.3\times10^7$ & $9.0\times10^{ 9}$ & $1.8\times10^{50}$ & $2.9\times10^{-1}$ & $3.5\times10^{-4}$ \\
      & 180 & 720 & 308 & 0.2 & $1.0\times10^8$ & $5.9\times10^{ 9}$ & $1.8\times10^{50}$ & $3.0\times10^{-1}$ & $4.1\times10^{-4}$ \\
\\
      & 720 & 432 &  77 &   1 & $9.8\times10^7$ & $1.0\times10^{10}$ & $1.8\times10^{50}$ & $2.9\times10^{-1}$ & $2.7\times10^{-4}$ \\
      & 720 & 288 &  77 &  10 & $4.3\times10^7$ & $8.7\times10^{ 9}$ & $1.7\times10^{50}$ & $2.7\times10^{-1}$ & $3.9\times10^{-4}$ \\
      & 720 & 100 &  77 & 100 & $1.1\times10^7$ & $7.4\times10^{ 9}$ & $1.6\times10^{50}$ & $2.4\times10^{-1}$ & $1.5\times10^{-4}$ \\
      & 720 &  36 &  77 & 500 & $3.8\times10^6$ & $6.5\times10^{ 9}$ & $8.7\times10^{49}$ & $1.3\times10^{-1}$ & $2.1\times10^{-4}$ \\
\hline
M6R4  & 720 & 720 &  77 & 0.2 & $3.9\times10^8$ & $1.3\times10^{10}$ & $3.6\times10^{50}$ & $3.9\times10^{-1}$ & $1.3\times10^{-3}$ \\
      & 360 & 720 & 154 & 0.2 & $2.9\times10^8$ & $1.2\times10^{10}$ & $3.7\times10^{50}$ & $4.0\times10^{-1}$ & $1.6\times10^{-3}$ \\
      & 180 & 720 & 308 & 0.2 & $2.2\times10^8$ & $1.1\times10^{10}$ & $4.0\times10^{50}$ & $4.2\times10^{-1}$ & $2.2\times10^{-3}$ \\
\\
      & 720 & 432 &  77 &   1 & $4.4\times10^8$ & $1.2\times10^{10}$ & $3.4\times10^{50}$ & $3.8\times10^{-1}$ & $1.0\times10^{-3}$ \\
      & 720 & 288 &  77 &  10 & $4.2\times10^8$ & $1.2\times10^{10}$ & $3.5\times10^{50}$ & $3.4\times10^{-1}$ & $4.3\times10^{-4}$ \\
      & 720 & 100 &  77 & 100 & $3.8\times10^8$ & $1.7\times10^{10}$ & $5.2\times10^{50}$ & $4.7\times10^{-1}$ & $2.7\times10^{-3}$ \\
      & 720 &  36 &  77 & 500 & $9.8\times10^7$ & $1.3\times10^{10}$ & $5.4\times10^{50}$ & $3.6\times10^{-1}$ & $6.9\times10^{-3}$ \\
\enddata
\end{deluxetable}

\subsection{C/O WD Results Summary}
\label{subsec:cowd}

\subsubsection{Peak Diagnostics}
\label{subsubsec:bulk_co}

Following the discussion from section \ref{subsec:hewd}, an equivalency can also be established
between the 2D C/O calculations and our previous 3D models.
Specifically, model M6R4 (M6R8) has identical initial isotopic composition
and a similar, though not exact, interaction strength as case B3M6R06 (B3M6R09) from \citet{Anninos18}.
Comparing entries from Table \ref{tab:peakdiag} against Table 2 of \citet{Anninos18},
we again find decent agreement between pairings for all diagnostics, providing further evidence
that these 2D models are proper stand-ins for elucidating 3D behavior
just as they were for the smaller mass He models. We remind the reader that
peak densities and temperatures are higher in the 2D models 
because they do not represent averages over extended domains as they do for the 3D calculations.
Also notice the ratio of calcium to iron production increases with increasing
perihelion radius or decreasing tidal strength, a behavior that is also observed in our 3D 
calculations and in the 2D and 3D helium star encounters.

\subsubsection{Initiation}
\label{subsubsec:initiation_co}

\begin{figure}
\includegraphics[width=0.9\textwidth]{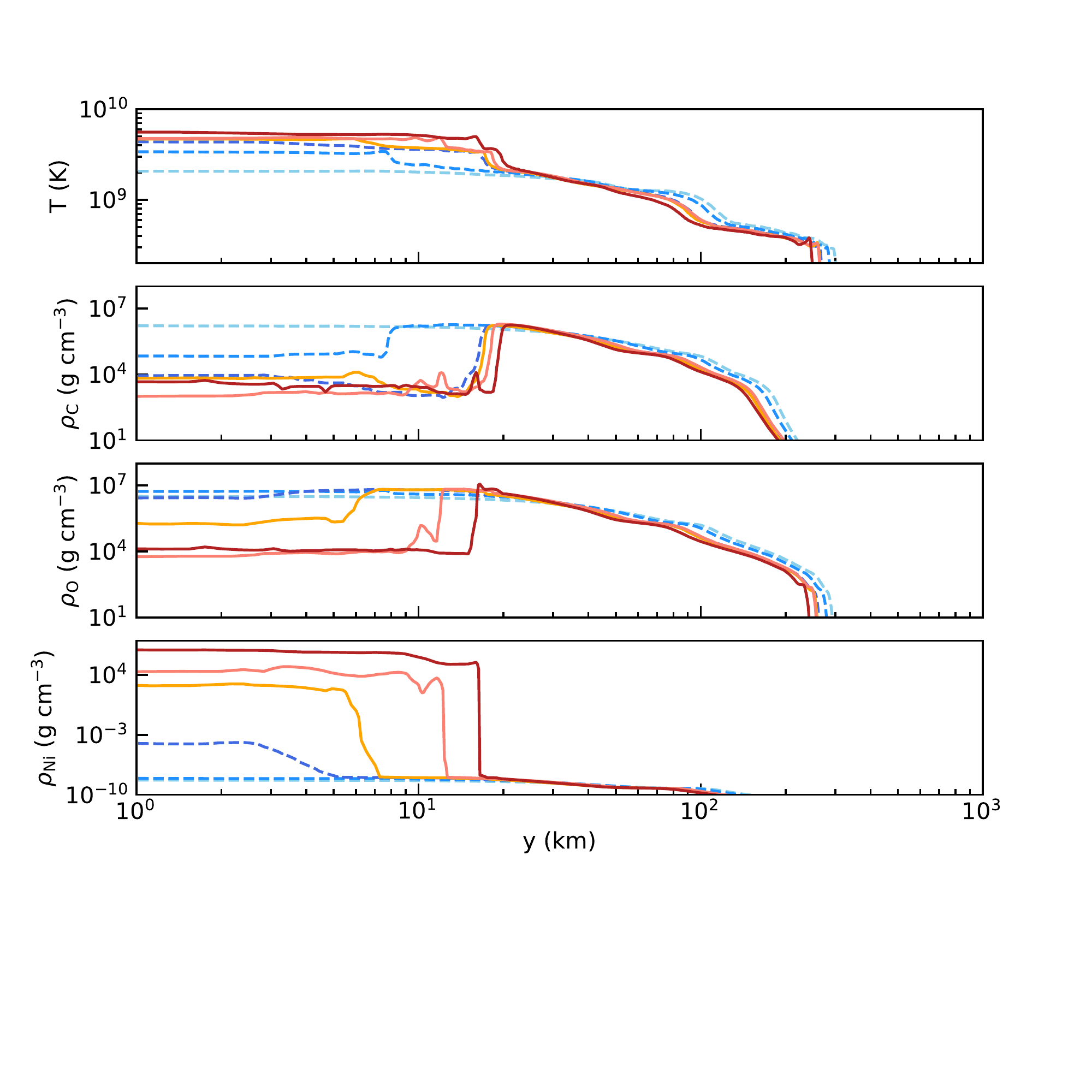}
\caption{
Line profiles showing the temporal evolution of temperature (top), carbon density, oxygen density, and nickel (bottom)
along the vertical $y$-axis for case M6R4.
Blue dashed (orange solid) lines correspond to solutions at different times before (after) sustained detonation producing nickel.
}
\label{fig:1D_rho_M6R4}
\end{figure}

Our closest C/O WD encounter M6R4 has a similar tidal strength ($\beta=9$) as the most distant He WD case
M2R12 ($\beta=8$), so they are interesting to compare given differences in composition and thermal support.
Figure \ref{fig:1D_rho_M6R4} plots the temperature, carbon density, oxygen density, and nickel in the manner 
of previous graphs, where dashed (solid) lines correspond to vertical profiles before (after) detonation,
ordered chronologically by tone of color (light to dark). For the most part results are similar to M2R12.
The temperature and total gas density first increase
adiabatically to conditions ripe for carbon burn, $\sim5\times10^6$ g cm$^{-3}$ and $\sim 2.5\times10^9$ K.
The reaction front during this early burn phase is traced by the mid-tone dashed blue curve in the carbon density plate of Figure \ref{fig:1D_rho_M6R4}.
Carbon burn raises the temperature behind the reaction front above $3\times10^9$ K while the gas density continues to increase
by accretion to about $10^7$ g cm$^{-3}$, promoting runaway burn to iron and nickel.
Notice the synchronous movement of the oxygen depletion contours and the nickel reactive fronts in the bottom two
plates of Figure \ref{fig:1D_rho_M6R4}.
There is a clear correlation between drops in oxygen levels with increases in nickel production,
as expected if the dominant species in the core is oxygen, not carbon. In fact we find that just prior to detonation
oxygen constitutes more than 60\% of the total mass behind the 10 km detonation site, while carbon represents less than 1\%. 
Most of the available fuel is converted promptly to nickel once detonation initiates following oxygen burn.
The total 20 km size hotspot (including above and below the plane) easily meets the minimum criteria of 1 to 2 km
\citep{Seitenzahl09} required to sustain detonation at these conditions. It is also two orders of magnitude greater than the spatial resolution.

There are signs in Figure \ref{fig:1D_rho_M6R4} that nuclear burn occurs violently enough
to disrupt the hydrodynamics. Notice for example the small local features imprinted on the density and temperature
plots inside 10 km. We observe similar features in profiles of the vertical velocity component. Close to the
orbital plane (less than a few kilometers) we even see evidence of a weak outflow where the velocity temporarily turns positive
before reversing itself. Although weaker than M2R12, the outflow correlates with the nickel production site, as it does for the He WD models, 
and is predominately a consequence of the energy released from the production of iron elements.

\begin{figure}
\includegraphics[width=0.7\textwidth]{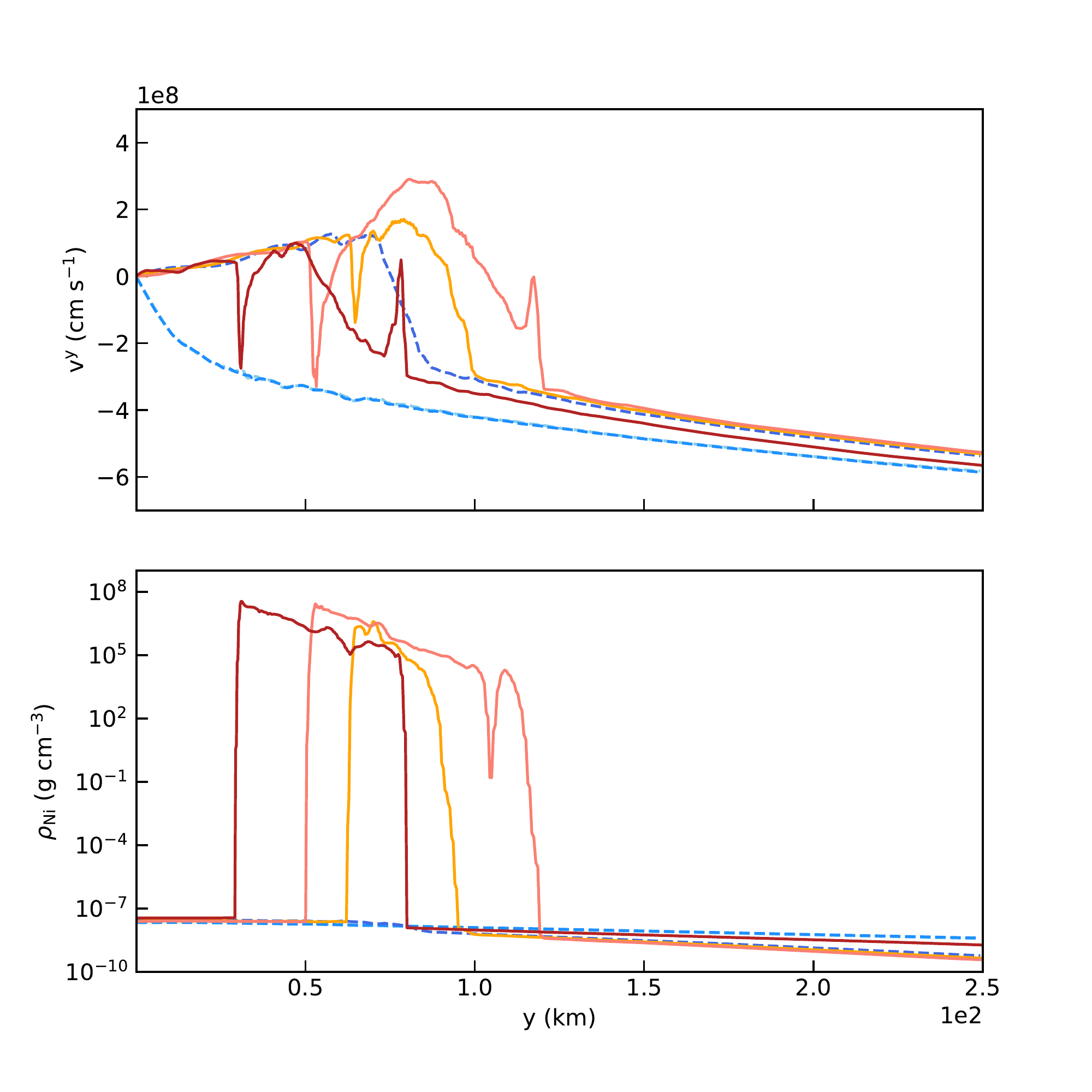}
\caption{
Line profiles of the vertical velocity (top plate) and nickel density (bottom) for case M6R8.
Dashed (solid) lines correspond to solutions at different times before (after) the detonation producing nickel.
}
\label{fig:1D_vel_M6R8}
\end{figure}

Our weakest interaction model M6R8 ($\beta=4$) behaves differently from any of the other cases.
Detonation triggers high off the orbital plane, at approximately 80 km (more than
an order of magnitude further out than M6R4),
and launches dual shock waves traveling in opposite directions, as shown in Figure \ref{fig:1D_vel_M6R8}, before the
outer shock eventually stalls and falls back towards the plane.
The  dark blue (dashed) curve in the top velocity plate is evidence of an outflow shock developing just before detonation
(notice the synchronization between nickel density fronts and shock positions, both of which are centered where the shock forms).
Detonation in this case is not initiated nor preconditioned by adiabatic compression from either homologous collapse or released energy:
Although the total gas density grows to $\gtrsim 10^7$ g cm$^{-3}$, the
temperature remains substantially colder than the other cases, barely exceeding $\sim 10^8$ K due to the weaker
tidal force that is unable to compress the star to the same internal pressures as M6R4. As a consequence
carbon and oxygen levels do not evolve but remain close to their initial steady-state composition ratios.
Additional collisional energy is necessary in this case to raise the adiabat and achieve robust detonation by what appears to be
direct initiation by the collision of gas streams moving in opposite directions.

\begin{figure}
\includegraphics[width=0.7\textwidth]{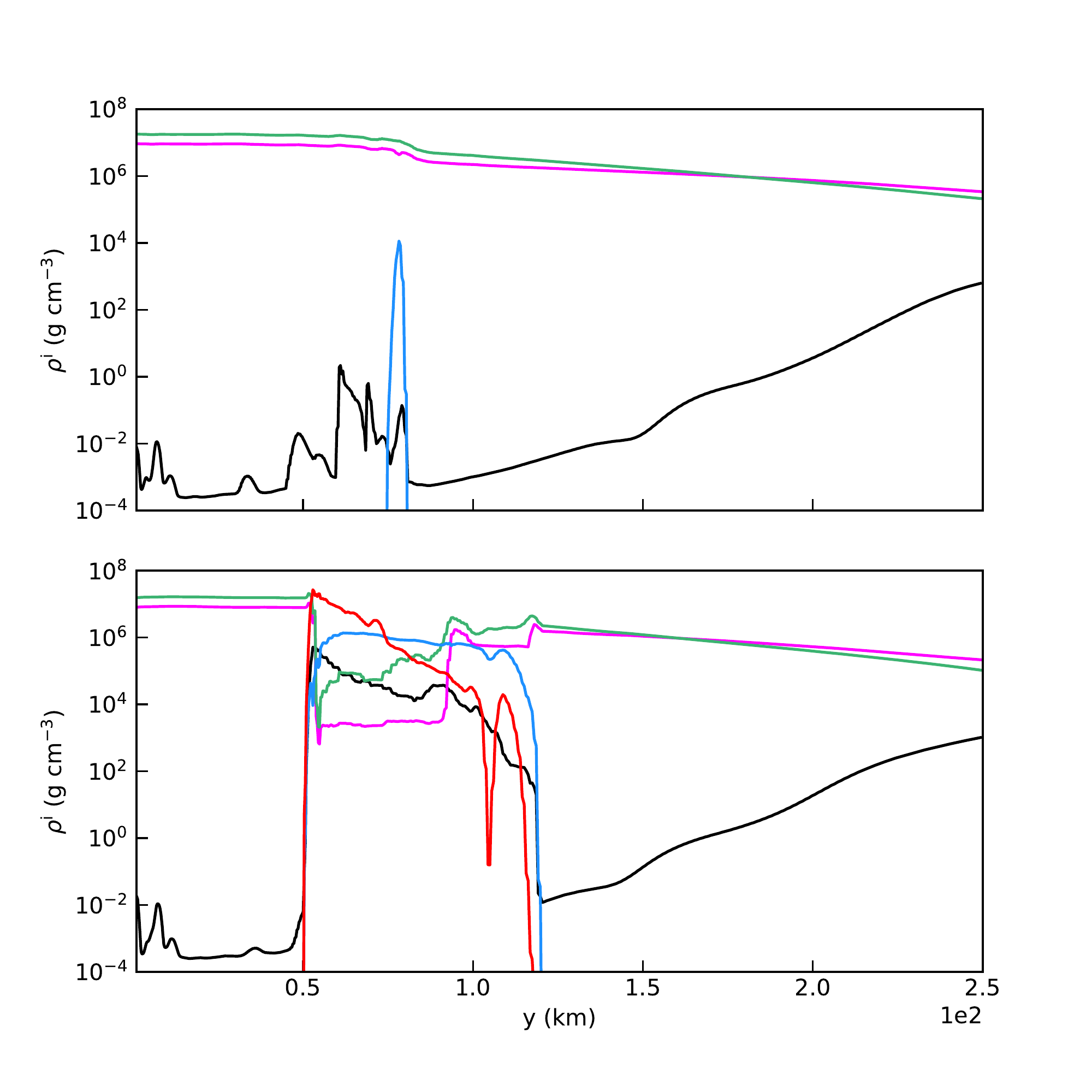}
\caption{
Line profiles of helium (black), carbon (magenta), oxygen (green), calcium (blue), and nickel (red) densities extracted
immediately before detonation (top) and immediately after (bottom) for M6R8.
}
\label{fig:1D_species_M6R4}
\end{figure}

Interestingly we find that even this weakest case scenario is capable of increasing temperatures in the interaction zones to
$5\times10^9$ K, where photodissociation of heavy to light elements becomes important. The reverse triple alpha reaction
in particular increases the helium mass fraction inside the hot detonation region up to a few percent, as observed
in Figure \ref{fig:1D_species_M6R4} where we plot the density distribution of atomic elements in the vicinity of
the initial hotspot. The top and bottom plates represent times just before and after detonation respectively,
corresponding to the dashed dark blue and solid yellow curves in Figure \ref{fig:1D_vel_M6R8}. We see first the creation of
calcium (blue) in a narrow region where the fluid streams collide (top plate), followed by robust burning 
of carbon (magenta) and oxygen (green) to nickel (red) as the detonation front expands in either
direction (bottom plate). The temperature behind the shock waves is hot enough to promote the photodissociation 
of heavy elements to helium, so that despite representing an insignificant part of the steady-state composition, helium
comes to make up a much greater fraction of the debris after disruption.

\subsubsection{Convergence Studies}
\label{subsubsec:convergence_co}

We performed the same series of horizontal and vertical resolution studies for model M6R4 as we had done for M2R4.
The results shown in Table \ref{tab:results_conv} are similar. We find little sensitivity in
either nuclear energy or iron group production to horizontal resolution, less than 10\%. But there is
significantly greater uncertainty with transient products (e.g., calcium) of nearly a factor of two.

Interestingly, we see an increase in nuclear energy production with decreasing vertical resolution, a trend
that is the reverse of what we found with low mass helium models. But in both models the errors
are fairly well constrained until resolutions coarsen to about 100 km cell size on the orbital plane. Although we don't observe
as dramatic a difference in iron group elements as we did for M2R4, this is
more than made up for with greater uncertainty in transient products which vary nearly ten fold over these scales.
A vertical cell size of 10 km again emerges as a reasonable compromise between accuracy and resource requirements.

\section{Conclusions}
\label{sec:conclusions}

Modeling the tidal disruption of white dwarf stars self-consistently with nuclear reactive networks
presents significant computational challenges, even with modern hardware and software advances.
Physical times range from microseconds (or less) to resolve nuclear physics,
and hours at minimum to carry that data out to the earliest phases of fall-back accretion.
The required spatial scales are just as demanding: from fractions of a kilometer
needed to capture nucleating hotspots, to a typical WD radius
of $10^4$ km, and an additional order of magnitude (or two) to
track the orbital trajectory. The literature is abound with examples of
different computational methods working towards this dynamical range in three spatial
dimensions, including SPH and adaptive Eulerian methods. However the question
as to whether or to what degree 3D calculations have achieved that goal, even for
the prompt nucleating phase which ends less than a second from periapsis, remains uncertain.

Some progress has been made recently
with 1D models undergoing (only) tidal compression. We have taken this opportunity to
expand on those 1D models and to develop a 2D analog for 3D behavior that mimics (both) tidal
compression and stretching in fewer spatial dimensions in order to increase
spatial resolution without stressing computational resources. We verified these 2D models
reproduce the thermodynamic state of compressed WD cores and the various phases of
disruption observed in 3D calculations, including nozzle flow, bounce-back, shock
wave generation, and the formation of detonation fronts. The 2D models produce comparable
density and temperature profiles and very similar global thermonuclear
yields and burn product distributions, scaling as they should with proximity to the BH.

We evaluated the effect of spatial resolution through a series of
convergence studies conducted along both horizontal and vertical
directions, parallel and perpendicular to the orbital plane respectively.
These studies were carried out for both He and C/O WD stars.
We find little difference in behavior or extracted diagnostics for horizontal
grid resolutions of up to $\Delta x_\mathrm{min} = 300$ km (our coarsest resolution)
for most of the diagnostics, including released nuclear
energy and iron group products. Intermediate mass elements exhibited some sensitivity to resolution,
but the scaling proved less than linear. 300 km grid resolution is not difficult to achieve, even in 3D.

Convergence studies along the vertical direction proved more interesting.
Although we found substantial differences in peak densities, the temperatures
were significantly less impacted, changing by less
than a factor of two across resolutions of $\Delta y_\mathrm{min} = 0.2$ to 500 km.
For diagnostics other than density, we find results converged to better than 10\% at 10 km resolution except again for
calcium production which remains uncertain to about a factor of two. Importantly, these results are in general
agreement with and support conclusions derived from earlier 1D models.

All of our calculations detonate and lead to sustained nuclear burn affecting a large fraction of
stellar material, with reaction fronts reaching in some cases to heights greater than 100 km off the orbital plane.
However the manner of detonation differs case to case, depending on the stellar mass
and composition (He vs C/O WD) as well as the tidal strength. 

Strong tidal forces ($\beta \sim 20$) in He WD models
develop and drive reaction fronts but also make it hard for shocks to keep pace, resulting in the separation
of shocks and burn fronts over time. Strong tidal forces and rapid pressure release from fast moving transverse
flows effectively freeze the shock in place
as the infalling gas continues to fuel thermonuclear burn and drive the reaction front further off the orbital plane.
These encounters initiate by homologous collapse and mild burn preconditioning in regions very close to the orbital plane.

Moderate strength encounters ($\beta\sim10$) between BHs and He or C/O WDs detonate in similar fashion,
by adiabatic compression and mild thermonuclear heating, but detonation is also facilitated by collisional heating
from outflows generated by thermonuclear energy released in the process of making nickel.
The latter effect appears to become increasingly important with decreasing $\beta$.
The tidal strength also dictates the location (distance from the orbital plane) where ignition triggers: 
larger $\beta$ encounters detonate closer to the plane.
In these cases the reactive fronts and shocks stay coincident after ignition, distinguishing them from
ultra-close encounters.

Weaker interactions ($\beta\sim4$) detonate even further off the symmetry plane, in isolated causally
disconnected regions. Ignition therefore does not occur simultaneously across the orbital plane,
but is propagated via oppositely directed shocks to the center from an initiation site many tens of kilometers away.
In these scenarios homologous collapse by tidal forces is not enough to heat the gas up to where it benefits from mild burn conditioning.
Detonation is instead triggered by reflected shocks.

In addition to differences in shock and reactive front behaviors, which we assume are representative of
a continuous spectrum of behaviors parameterized by $\beta$, we also observe interesting differences in the
early pre-detonation phase. Most of the cases studied here generate
significant levels of not only intermediate mass products (e.g., calcium) but also of nickel elements before the onset of detonation.
This early and relatively mild build-up of nuclear ash occurs adiabatically and can in some cases produce transient outflows, depending
on the interaction strength and WD model. The extent to which these outflows reach before detonation triggers depends inversely on tidal strength.
Sustained detonation does eventually set in for most cases after having already produced significant quantities of heavy elements adiabatically,
including nickel. The exception is the weakest interaction example which does not heat enough to support even mild preconditioning.
In this case the tidal stream maintains a relatively cold temperature of $\gtrsim2\times10^8$ K until a reflected shock
forms and launches an outflow, raising the gas adiabat to detonation conditions by collisional heating.

\added{
In summary, the combination of adiabatic compression aided by mild carbon/oxygen preconditioning are the main nickel-producing detonation
mechanisms at intermediate to high tidal strengths. Reflected shocks or collisional outflows represent a third initiation
process that becomes increasingly important with weaker, more distant encounters. They can be the dominant
physical interactions driving  initiation at the lower end of tidal strengths capable of perpetuating nuclear reactions.
}

\begin{acknowledgments}

This work was performed in part under the auspices of the U.S. Department of Energy by 
Lawrence Livermore National Laboratory under Contract DE-AC52-07NA27344. Some of the computing for this project was performed on the Beocat Research Cluster at Kansas State University, which is funded in part by NSF grants CNS-1006860, EPS-1006860, EPS-0919443, ACI-1440548, CHE-1726332, and NIH P20GM113109.

\end{acknowledgments}

\software{ Cosmos++ \citep{Anninos05, Anninos12, Anninos17}, MESA \citep{Paxton11}, Torch \citep{Timmes99} }

\listofchanges
\end{document}